\pdfoutput=1
\documentclass{aamas2017}

\usepackage{times}
\usepackage{helvet}
\usepackage{courier}
\usepackage{graphicx}

\newcommand{\transp}{\top}

\newcommand{\mytilde}{\raise.17ex\hbox{$\scriptstyle\sim$}}

\usepackage[boxed, noend, noline]{algorithm2e}
\usepackage{amsmath, amssymb}
\begin{document}
\sloppy
\hyphenation{Abstractions}
%\hyphenation{Overhead}

\title{Context-Based Concurrent Experience Sharing in Multiagent Systems}

\numberofauthors{4}

\author{
% 1st. author
\alignauthor
Dan Garant\\%\titlenote{Dan Garant}\\
      %\affaddr{College of Information and Computer Sciences}\\
      \affaddr{University of Massachusetts}
	  \affaddr{Amherst, MA 01002}\\
      \email{dgarant@cs.umass.edu}
%2nd. author
\alignauthor
Bruno C. da Silva\\%\titlenote{The secretary disavows any knowledge of this author's actions.}\\
      %\affaddr{Institute of Informatics}\\
      \affaddr{Federal University of Rio Grande do Sul. Porto Alegre, Brazil 91540-000}\\
      \email{bsilva@inf.ufrgs.br}
%3rd. author
\alignauthor
Victor Lesser\\%\titlenote{Dan Garant}\\
      %\affaddr{College of Information and Computer Sciences}\\
      \affaddr{University of Massachusetts}
	  \affaddr{Amherst, MA 01002}\\
      \email{lesser@cs.umass.edu}
%4th author
\and
\alignauthor
Chongjie Zhang\\%\titlenote{Dan Garant}\\
      %\affaddr{CSAIL}\\
	  \affaddr{Tsinghua University}\\
      \affaddr{Beijing, China}\\
      \email{chongjie@tsinghua.edu.cn}
}

\maketitle

%----- ACTUAL PAPER STARTS HERE ------------------------------
\begin{abstract}
One of the key challenges for multi-agent learning is scalability. In this paper, we introduce a technique for speeding up multi-agent learning by exploiting concurrent and incremental experience sharing. This solution adaptively identifies opportunities to transfer experiences between agents and allows for the rapid acquisition of appropriate policies in large-scale, stochastic, homogeneous multi-agent systems. We introduce an online, distributed, supervisor-directed transfer technique for constructing high-level characterizations of an agent's dynamic learning environment---called contexts---which are used to identify groups of agents operating under approximately similar dynamics within a short temporal window. A set of supervisory agents computes contextual information for groups of subordinate agents, thereby identifying candidates for experience sharing. Our method uses a tiered architecture to propagate, with low communication overhead, state, action, and reward data amongst the members of each dynamically-identified information-sharing group. We applied this method to a large-scale distributed task allocation problem with hundreds of information-sharing agents operating in an unknown, non-stationary environment. We demonstrate that our approach\footnote{A more complete presentation of our approach, as well as additional experiments, can be found in \cite{garant2015accelerating}.} results in significant performance gains, that it is robust to noise-corrupted or suboptimal context features, and that communication costs scale linearly with the supervisor-to-subordinate ratio.
% discuss a favorable trade-off between computational complexity, communication overhead, and performance.
\end{abstract}

%----- KEYWORDS SECTION --------------------------------------
% Note that the category section should be completed after reference to the ACM Computing Classification Scheme available at
% http://www.acm.org/publications/class-2012
% Hint: a useful place to start could be  Computing methodologies →  Artificial intelligence →  Distributed artificial intelligence
% The block below is an example generated by the ACM web page. Replace it with appropriate block for your work.
% <concept_id>10010147.10010178.10010219.10010220</concept_id>
% <concept_desc>Computing methodologies~Multi-agent systems</concept_desc>
% \ccsdesc[500]{Computing methodologies~Multi-agent systems}
% % end auto-generated block
% \printccsdesc
% \keywords{Experience Sharing, Multiagent Systems, Distributed Task Allocation}
%--------------------------------------------------------------

\section{Introduction}
\label{sec:intro}

In large-scale multi-agent systems consisting of hundreds to thousands of reinforcement-learning agents, convergence to a near-optimal joint policy, when possible, may require a large number of samples. These systems, however, may contain groups of agents working on nearly identical local tasks or under approximately similar environmental dynamics. Identifying such groups may prove useful in cooperative domains, due to the opportunity of exploiting shared information. Information sharing has been extensively studied in single-agent settings with the goal of transferring knowledge from a source task to novel tasks \cite{taylor2009transfer,lazaric2008transfer,carroll2005task}. Applying this idea to the multi-agent setting (MAS), it is apparent that experiences may be transferred not only across similar tasks, but also between concurrently-learning agents in a shared environment. This paper focuses on the problem of online transfer of experiences between such agents---with an emphasis on the adaptive discovery of groups of agents where experience sharing is possible and beneficial.

\begin{figure}[!ht]
\centering
\includegraphics[width=0.98\linewidth]{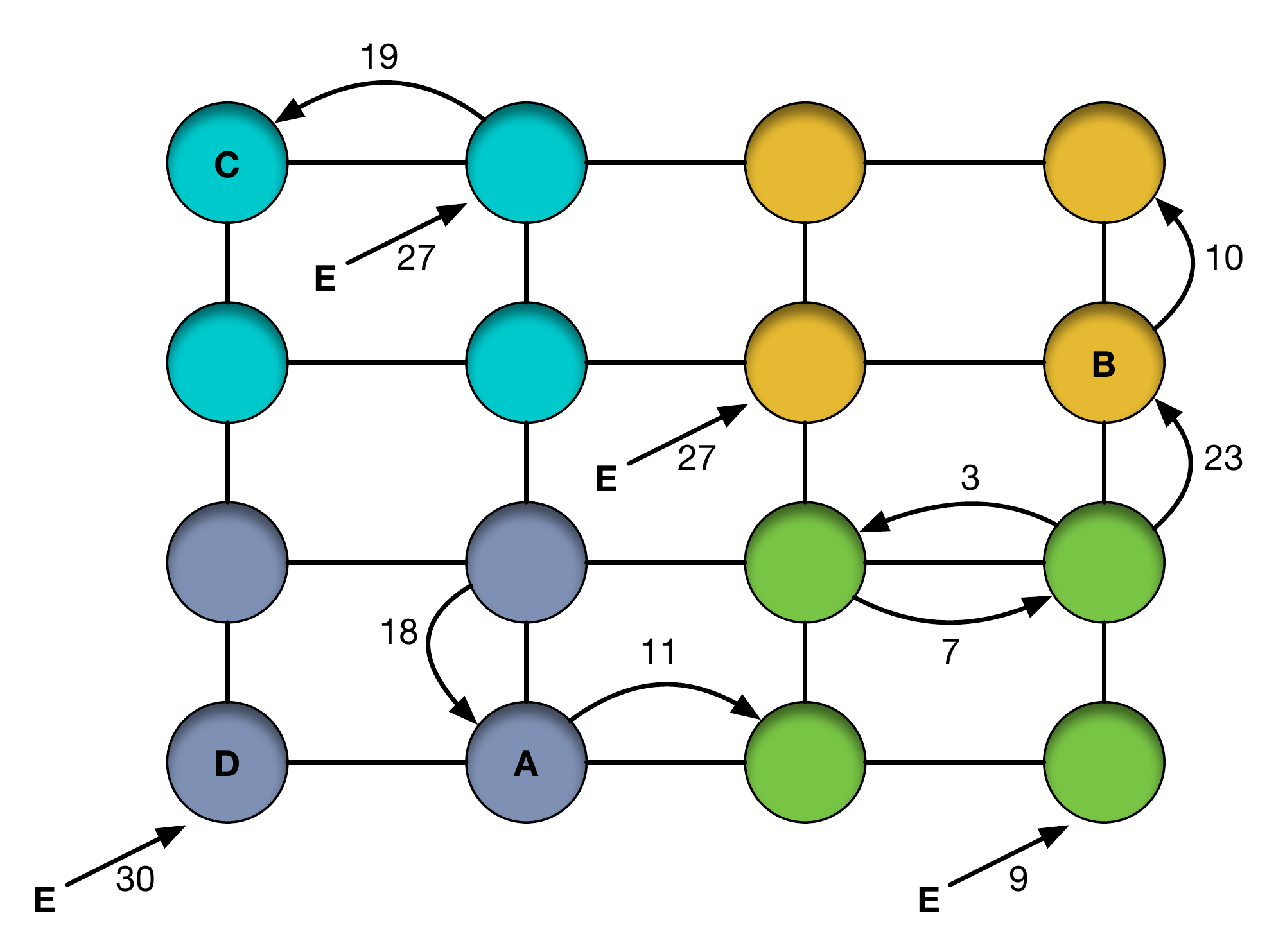}
\caption{Sample task allocation network.} 
\label{fig:task-allocation-example}
\end{figure}

In multi-agent settings, agents need to interact and learn concurrently. The environment, from each agent's perspective, is non-stationary due to the presence of other concurrently-adapting agents. Since the observations made by one agent are conditioned on the behaviors of its neighbors, it is not clear when they can be usefully exchanged and reused by other agents---which may be operating under different local environments and may be interacting with different types of neighbors. Experience exchange is, therefore, not straightforward in non-stationary MAS. As an example, consider the task allocation problem depicted in Figure~\ref{fig:task-allocation-example}. 
 % TODO: what are node colors?
Agents are represented as nodes and may receive tasks from the environment or from other agents. Each agent can choose to fulfill a given task or to forward it to a neighboring agent. Assume that agents have partial knowledge of the system: they do not know the global structure of the network nor have access to state or policy information of other agents. This results in a non-stationary problem where it may be inappropriate to transfer information between some pairs of agents. Agent D, for instance, receives a large number of tasks from the environment and may need to forward them to a neighbor; agent C receives tasks from a neighbor and may need to direct them away from its heavily-loaded neighbor. Agents A and B, on the other hand, undergo similar task-forwarding patterns with respect to their neighbors. Experience transfer, then, may be appropriate between agents A and B (said to be \emph{contextually compatible} agents), but not between agents D and C.

To address the information-sharing problem in non-stationary MAS we propose modeling \emph{contexts} as inherently dynamic local characterizations of the environment under which agents operate. They are defined over short timescales during which policies and models are approximately static. In Sections \ref{sec:context-features} and \ref{sec:context-based-learning} we introduce and motivate a context-similarity measure grounded in the comparison of abstract representations of environment dynamics, rather than policies or Q-values, and advocate the use of \textit{supervisory} agents (which periodically collect data from subordinates) as a way of identifying contextually-compatible agent groups where experiences may be shared.
As will be further discussed in the following sections, contextual modeling is made possible through the commonly-studied property of interaction sparsity \cite{witwicki2010influence}, or what Simon \cite{simon1996sciences} referred to as \emph{nearly-decomposable systems}. This is apparent in many domains, such as distributed task allocation, disaster planning \cite{kitano1999robocup,oliehoek2008exploiting} and sensing networks \cite{zhang2013coordinating}, in which agents interact strongly with only a small group of closely-related partners.% (defined according to task-based or proximity metrics). 

Although other transfer mechanisms are possible, to our knowledge no other methods exist that address the particular setting and scale presented in this paper. We believe this is the first algorithm that allows experience sharing in a concurrent and interacting MAS with \mytilde1000 agents while undergoing low communication and computational overhead. We evaluate our method on a large-scale distributed problem and demonstrate that context-based transfer yields significant performance gains. We further show \emph{1)} that the time complexity of our method scales with the number of agents within each supervisory group, not with the total number of agents in the network; \emph{2)} that our method is robust to noise-corrupted or suboptimal context features; \emph{3)} that communication costs scale linearly with the supervisor-to-subordinate ratio; and \emph{4)} that sparse lossy compression schemes may be deployed and provide significant improvements in communication costs, while inducing negligible negative impact on system-wide performance.

%We also show that our method is robust to noise-corrupted or suboptimal context features, that communication costs scale linearly with the supervisor-to-subordinate ratio, and empirically demonstrating that even when simple (and possibly corrupted) context characterizations are used, experience sharing opportunities are still identified and exploited.

\section{Related Work}

In this section we discuss related work that also aims at experience sharing in MAS. Kretchmar et al. introduced a technique for agents to periodically exchange Q-values to accelerate learning \cite{kretchmar2002}. They assumed, however, that agents operate on independent copies of an environment and do not interact. Boutsioukis et al. relaxed this assumption via a method for using Q-values of a source task to bias the initial policy of target tasks \cite{boutsioukis2012transfer}. They assumed that learning on the source task had to be completed before transfer was made possible, and required the use of inter-task mappings. We do not assume that such mappings are needed and instead infer when observations may be transfered by identifying groups of agents that operate under similar local contexts. Taylor et al. introduced a transfer method that allowed for source and target tasks to be learned in parallel \cite{taylor2013transfer}. It implicitly assumed that all agents experienced tasks with similar state values---which may not be true if they operate in contexts with different transition dynamics. More recently, Mnih et al. introduced a technique for accelerating deep learning algorithms via asynchronous sharing of policy gradients \cite{mnih2016}. This allowed for independent agents to cooperate in solving a complex task, but required that agents did not interact while doing so. In Section \ref{sec:context-features} we extend the discussion presented here and introduce related techniques relevant to the problem of characterizing local contexts in order to identify sharing opportunities.

\section{Setting}

% \begin{align*}
% R_i(r_i | S, a_1, a_2, \ldots, a_n),
% \end{align*}

Multi-agent decision-making problems are often framed in the context of Markov games \cite{owen1982}. Markov games model $n$ agents operating in an environment described by a joint state $S$. A state transition function specifies the conditional probability of the environment transitioning to state $S'$, given that it was in $S$ and that agents executed a particular joint action $(a_1, \ldots, a_n)$; i.e., $P(S' | S, a_1, \ldots, a_n)$. In Markov games, each agent $i$ holds a particular reward function $R_i(r_i | S, a_1, \ldots, a_n)$, which we consider here to be a conditional distribution over rewards.

In cooperative environments, individual reward functions may be identical---each agent's individual performance perfectly aligns with the system's performance. In this paper we consider the more general case of \textit{decomposable} reward functions, which arise in structured settings such as Network Distributed-Partially Observable Markov decision processes \cite{nair2005networked} or factored multi-agent MDPs \cite{guestrin2001multiagent}. We also assume that the global state $S$ is decomposable into (potentially overlapping) components $s_i$, each of which represents the portions of the state that agent $i$ can directly observe. This arises in systems where it may be infeasible for agents to learn over the full joint space or when network structure or communication bandwidth introduce limitations on state observability. In general, the state observable by individual agents may be insufficient to faithfully reconstruct the overall state transition model, $P_i(s_i' | S, a_1, a_2, \ldots, a_n)$. Motivated by the idea of interaction sparsity \cite{witwicki2010influence}, we address this difficulty by observing that sparsity in $P_i$ may allow observations collected from a small number of neighboring agents to be used in order to reasonably estimate $P_i$. 

%Applied to the running example in Figure~\ref{fig:task-allocation-example}, this means that the global state $S$ (corresponding to the queue length of all agents) is not observable by all agents. Each agent $i$ can only observe $S_i$---its own task load. Furthermore, agents have no access to the global reward signal, defined here as the overall average task service time of all agents, and only observe the actions of its direct neighbors. These conditions and constraints induce non-stationarity in the system's state transition model, resulting in a non-trivial learning process.

\section{Overview of the Method}

Before introducing the technical details of the method we propose, we start by presenting a high-level summary of the steps involved in determining sharing opportunities by grouping agents based on their local learning environments (or \emph{contexts}):

\begin{enumerate}
    \item each agent collects observations from its local environment in the form of state, action, reward, and next state tuples. Every $K$ time steps (the \emph{reporting interval}), agents report such observations to their corresponding supervisors;
    \item supervisors use the received information and their local knowledge about the interactions between subordinate agents to compute \emph{context summary vectors}, one per agent. These vectors correspond to dynamic local characterizations of the environment under which agents operate, and are used to identify possible sharing experiences; %the local environment that each agent operates on;	
    \item supervisors measure the similarity between the context summary of each subordinate agent with respect to a covariance-appropriate and scale-independent metric; similar agents are organized into sharing groups;
    \item supervisors relay experiences (state, action, reward, next state tuples) between all members of each sharing group;
    \item return to step (1) and adaptively regroup agents according to updated context information.
\end{enumerate}

\begin{figure}[!h]
\centering
\includegraphics[width=\linewidth]{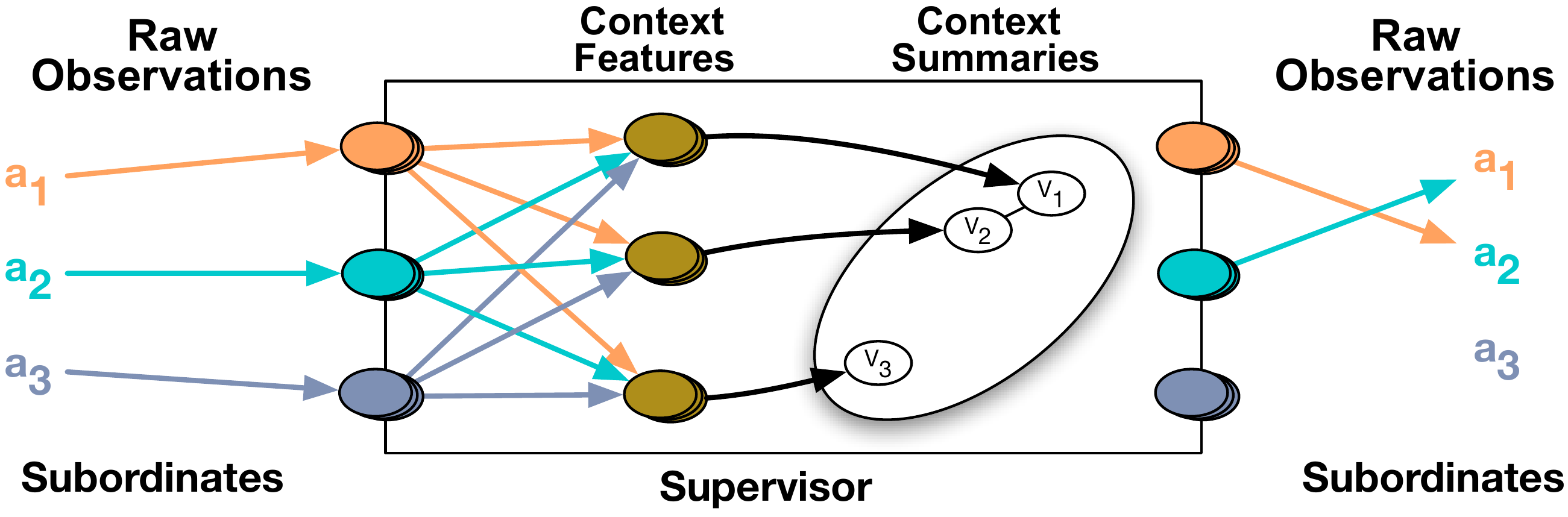}
\caption{Overview of the context-based transfer process.}
\label{fig:context-construction}
\end{figure}

Intuitively, a supervisor periodically collects information from a small number of subordinate agents in its \emph{supervisory group} and computes context features. These are embedded in a summary space in which similarity stochastically determines \emph{sharing opportunities}---not all agents within a same supervisory group need to share experiences. Note that the method we propose here does not aim at finding optimal subordinate-supervisor assignments, but on efficiently identifying sharing opportunities within a given supervisory structure. Sharing opportunities between agents are dynamically re-evaluated by our method based on updated information collected in a reporting interval. The overall context-creation and data-transfer process is depicted in Figure~\ref{fig:context-construction}. 

\section{Context Features as State Abstractions}
\label{sec:context-features}

Context features are compact abstractions of the local learning environment under which an agent operates. Carrol \& Seppi \cite{carroll2005task} discuss the difficulty of constructing such abstractions given the difficulty of defining similarity metrics between learning environments. In the single-agent RL setting, many metrics are possible; most compare local environments by comparing policies, Q-values, or reward function differences \cite{boutsioukis2012transfer,hu2015learning,taylor2013transfer}. In the multi-agent setting, however, we wish to capture a measure of \emph{compatibility} in the local learning environment of agents. 

Compatibility-based metrics are helpful to avoid issues arising by using policy, Q-function or model similarity as proxies of environment similarity. Consider, for instance, the problem of experience imbalance: the policies or Q-values under comparison need to be constructed with enough samples so that they are accurate estimates of an optimal policy \cite{price2003accelerating}. Metrics based on policy comparison are also difficult to define since optimal policies for Markov games are not unique \cite{littman2001value}, thus making policy similarity a poor indication of environment similarity. Metrics based on Q-function comparison are also non-trivial to define since latent features often cause agents to operate in different state spaces. %A fourth problem affecting these metric is due to the possibility that the agents being compared visit disjoint regions of the state space. In this case, their Q-values, policies, and reward models could be well-estimated in separate regions of the space, making a meaningful direct comparison difficult. 

To avoid these problems, we propose reasoning over the underlying latent model of a stochastic game via \emph{contextual comparison}. In particular, if we can identify that agents are working under a same local transition and reward model, we can infer (from the homogeneity of the system) that they are facing the same learning problem. Experiences gathered by each compatible agent are interchangeable and can be transferred. Since estimating system-wide transition and reward models is impractical in large systems, we rely on the use of context features to form broad-scope summaries, or \emph{abstractions}, of transition and reward models as experienced by individual agents. Abstractions in RL have been studied extensively, and are not the focus of this paper; we simply rely on any of the many existing methods (e.g. \cite{dietterich2000,balaraman2003,mahadevan2005,keller2006,li2006,paar2007,konidaris2009}) to construct features capable of abstracting the state of a particular problem at hand. If prior domain knowledge is available, context features may also be manually specified to abstract an agent's state variables---and possibly those of its observable neighbors. The way in which sufficient statistics and features for such abstractions are defined and computed depends on the structure of the MAS and needs to be defined in light of the characteristics of the application at hand. In this paper we empirically show that even simple (and possibly noise-corrupted) abstractions are often sufficient to allow for experience sharing opportunities to be identified in large-scale non-stationary systems composed of hundreds of concurrently-learning agents.

%\footnote{In systems with few agent interactions, e.g., an agent's local state may be sufficient to characterize its local learning environment. In more complex non-stationary systems, context features need to account for agent interactions. If these are sparse, an agent's transition model is influenced by few other agents and the non-stationarity affecting it is bounded, and a model may be well-approximated from experiences collected in a small neighborhood.} 

\section{Context-Based Learning}
\label{sec:context-based-learning}

As mentioned above, \emph{context features} are abstractions of the local learning environment of an agent, and are constructed based on data collected at a particular time $t$. In order to characterize the broader context of an agent's learning environment---for instance, the medium-term effects of its interactions with neighboring agents---one needs to combine context observations across a time window\footnote{As will be discussed later, selecting a time window has implications on runtime, communication overhead, and reliability. See Section~\ref{sec:results} for more details.}. We refer to this aggregate context information as a \emph{context summary vector}. Context summaries are computed by having neighborhoods of agents (called \emph{supervisory groups}) send local context information to their common supervisor. The supervisor annotates these with information about the states and actions experienced by the agents it oversees. The contextual information received by the supervisor is then used to compute a context summary according to the method described in Section \ref{subsection:context-similarity} and Algorithm \ref{alg:sharing-partners}.

When comparing context summaries to identify possible sharing experiences, it is necessary to select an appropriate distance metric in context space. This metric depends on the distribution from which context features are drawn. Unless a designer has \emph{a priori} information about this distribution, we assume that it can be approximated by a multivariate Gaussian. This is justified by two reasons: \emph{1)} by the central limit theorem (CLT), applied here in the limit of $K$;\footnote{A variant of the CLT for weakly dependent processes can also be applied assuming that sufficiently separated agents have approximately independent experiences \cite{de2000functional}.} and \emph{2)} because this is the distribution that imposes minimal prior structural constraints (it is the maximum-entropy distribution for these parameters) in the absence of prior knowledge. From this assumption, it follows that context features can be summarized into a context summary via a \emph{mean context vector and its covariance matrix}, and that a natural scale-invariant distance metric to compare context summaries exists: the Mahalanobis distance~\cite{murphy2012machine}. 

Mahalanobis metrics generalize Euclidean distances in a way that naturally takes correlations of the dataset (i.e., correlations between context features of different agents) into account. Furthermore, these distances are preserved under full-rank linear transformations of the space spanned by the data. This implies that context distances are preserved even if the context features are further abstracted or transformed under non-degenerate projections down onto any other context space. Even though this metric is quite general (it is naturally data-adaptive, scale-invariant, and preserved under non-degenerate transformations) and imposes the least prior structural constraints in the absence of expert knowledge, other metrics may be used. In a designer chooses to do so, our method changes trivially---the Gram matrix used to stochastically identify sharing opportunities (see Algorithm \ref{alg:sharing-partners}) is in that case computed according to the alternative selected metric.

\subsection{Agent Organization}
\label{subsec:agent-organization}

Agents are dynamically organized by our method in \emph{sharing groups} whenever they are close (in context space) to other agents within a supervisory group. This organization process involves a trade-off between how many supervisors exist in the system and the number of agents within each sharing group. As the number of supervisors grows, the sharing assignment problem becomes increasingly distributed, reducing the computational requirements imposed on any individual supervisor. On the other hand, supervisors overseeing larger groups of subordinates are capable of selecting from a larger pool of experiences, increasing the likelihood that similar agent groups can be identified. Each supervisor is generally responsible for a set of subordinates selected through self-organization~\cite{zhang2010self} or directly given the network structure. In this work we do not focus on finding optimal subordinate-supervisor assignments, but on efficiently identifying sharing opportunities within a given supervisory structure. In the experiments presented in Section \ref{sec:results} we evaluate our method in a network of hundreds of agents cooperatively solving a large-scale distributed task allocation problem; in this case, subordinate-supervisor assignments are determined in a way that supervisors span physical regions of the network consistent with agent interaction strength \cite{zhang2010self}. Such an agent-organization criterion is justified in this task due to the assumption of interaction sparsity \cite{witwicki2010influence}, a general characterist of nearly-decomposable systems. Many other real-world multi-agent systems with similar local sparse network-like interactions exist and could be organized similarly---ranging from disaster planning systems \cite{kitano1999robocup,oliehoek2008exploiting} to sensing networks \cite{zhang2013coordinating}.

%\footnote{A guideline is to distribute supervisors to effectively span a physical space or in a way consistent with agent interaction strength \cite{zhang2010self}.}. In this work we focus our efforts in developing a method for effectively identifying sharing experiences that do not require excessive communication bandwidth.

\subsection{Assessing Context Similarity}
\label{subsection:context-similarity}

We now present a method for computing contextual similarity between agents and forming sharing groups. Suppose that an agent communicates $K$ observations to its supervisor every $K$ time units. Our solution easily extends to cases where agents do not make one observation per time step. Let $O_i$ be a time-indexed experience vector of agent $i$:
\[
O_i = [ (s, a, s', r)_{t_1}, (s, a, s', r)_{t_2}, \ldots, (s, a, s', r)_{t_K} ]^\transp.
\] 
A supervisor overseeing $n$ agents computes contextual information by mapping $\Omega = \{O_1, \ldots, O_n\}$ into an $n$-tuple of \emph{context summary vectors} $V = (V_1, \ldots V_n)$. Assume we are given a function $f$ for computing \emph{context features} for agent $i$ at time $t$, given the history of observations $\Omega^t = \{ O_1^t, \ldots, O_n^t \}$, where $O_i^t = \{ (s, a, s', r)_h | h \leq t \}$. That is, $\Omega^t$ contains the observations of all $n$ agents in the neighborhood up to some time $t$. The context features for agent $i$ at some time $t \in [t_1,\ldots,t_K]$ are given by $f(i, \Omega^t)$. Note that $f$ may use information about neighboring agents when constructing features that describe $i$'s local learning environment. This yields a total of $nK$ context features vectors per supervisor. Each context feature vector is a sample drawn at a particular time from the (latent, unknown) underlying context distribution of an agent. These samples can be combined by the supervisor to compute an unbiased estimate of the true mean of the underlying context distribution. Unbiased estimates of the true mean vector of the context distribution are called \emph{context summary vectors}, and are compact descriptions of the learning environment of each agent within a supervisory group. The supervisor stores the context summary vectors of its $n$ subordinates in a tuple $V$:

\[
V = \Bigg( \frac{1}{K} \sum_{t=t_1}^{t_K} f(1, \Omega^t), \ldots, \frac{1}{K} \sum_{t=t_1}^{t_K} f(n, \Omega^t) \Bigg)
\]

% As described earlier, automated RL abstraction methods may be used to construct context features $f$. In the task allocation domain, this is not needed since meaningful context features can be directly derived from the states and actions of neighboring agents, and correspond to straightforward measurements of the local neighborhood surrounding an agent (). They encode the rates with which an agent's neighbors receive tasks from the environment and from other agents, and also the relationship between the load (queue length) of an agent and its neighbors.
% $\gets \text{Bernoulli}\left(\frac{\exp \left(M^i_{h,j} \right)}{  \sum_{a, b \in C_i} \exp \left( M^i_{a, b} \right)}\right)$\;

Note that each element of $V$, as computed above, is an unbiased estimate of the true mean context vector under the distribution assumptions made in Section \ref{sec:context-based-learning}. If a different distance metric is selected by a domain expert, the elements of $V$ need to be defined so that they correspond to unbiased estimators of the mean of the corresponding underlying distribution posited by the designer.

Our method for identifying sharing opportunities is based on a stochastic sampling process that probabilistically partitions agents into sharing groups, based on their contextual similarity. In particular, membership of an agent to a sharing group is stochastically determined based on the similarity of that agent's context summary and the context summaries of other agents in the same sharing group. This stochastic process partitions subordinate agents within a given supervisory group so that agents operating under similar underlying local dynamics have a higher probability of undergoing experience sharing.

The sampling process that we define consists in a two-stage selection routine. First, agents are partitioned into \emph{potential sharing groups} $C_1, \ldots, C_k$, based on to their similarity. Here, similarity is measured with respect to the selected context distance metric. Potential sharing groups are sets of agents (within a same supervisory group) that, given their contextual similarity, are deemed to be feasible candidates to undergo experience sharing. Partitioning agents into potential sharing groups is an unsupervised process that can be implemented via any standard clustering algorithm. Its purpose is to ensure that the computational costs of stochastically sampling agents in order to construct sharing groups is approximately constant, independently of the number of agents within a supervisory group. In particular, it ensures that the Gram matrix used to define the sampling distribution (see next paragraph and Algorithm \ref{alg:sharing-partners}) has dimensions that scale linearly with $k$. We say that agent $A_i \in C_j$ if that agent's context summary $V_i$ belongs to potential sharing group $C_j$. 

Next, the tuple $V$ of context summaries collected by a supervisor is used to define a probability distribution that stochastically determines whether particular pairs of subordinate agents should belong to a same sharing group. First, pairwise distances are computed over the context summaries of every pair of agents $A_h$ and $A_j$ in a potential sharing group $C_i$. This distance is stored in the $(h,j)$-th entry of a Gram matrix $M$. Each $h$-th row of $M$ corresponds, therefore, to distances in context space between an agent $A_h$ and all other agents in a potential sharing group $C_i$. We use $M$ to define a sampling distribution $P_i$ that stochastically determines whether pairs of subordinate agents in $C_i$ should belong to a same sharing group, given their contextual similarities. In particular, $P_i(h,j)$ denotes the probability that any two agents $h$ and $j$ within $C_i$ will be assigned to a same sharing group, based on their distances in context space. In this paper we define $P_i$ as a Boltzmann distribution constructed based on the pairwise context summary distances between agents in $C_i$: 
\[
P_i(h,j) \equiv \frac{\exp \left(M^i_{h,j} \right)}{\sum_{a, b \in C_i} \exp \left( M^i_{a, b}\right)}
\]
Boltzmann distributions are widely used in machine learning when one needs to define probability distributions that depend on the relative difference between numerical quantities associated with each element in a given population. Here, they depend on the distance between context summaries of any given pair of agents in $C_i$. Note that $P_i$, as defined above, assigns a probability to every pair of agents in $C_i$ and reflects how likely it is that those agents will be selected for membership in a same sharing group. Agents are selected for membership in a sharing group by sampling from $C_i$ without replacement; this ensures that agents will belong to at most a single sharing group---see Algorithm \ref{alg:sharing-partners} for details. This selection process is repeated in order to construct a group-sharing function $\Psi$, which maps agents to sharing groups. Once $\Psi$ has been established, supervisors relay all observations within the $K$-unit time window from all agents in $\Psi(i)$ to agent $i$. Agents incorporate these experiences into their policies using any off-policy learning algorithm. Note that an agent $i$ within a potential sharing group $C_i$ is not necessarily associated to any sharing partners; if $i$ is dissimilar from all other $(n-1)$ subordinates, $\Psi(i) = \varnothing$ with high probability.

\begin{algorithm}
\DontPrintSemicolon
%\SetNoline

\KwIn{Set of agents $A=\{1, 2, \ldots, n\}$}
\KwIn{Tuple $V = (V_1, \ldots, V_n)$ of context summaries}
\KwOut{Mapping $\Psi:A \to \mathcal{P}(A)$ from agents to sharing groups \footnotesize{\,\,($\mathcal{P}(\cdot)$ \emph{denotes powerset})}}

Let $\mathcal{M}$ be a selected context distance metric. \;
Partition $V$ into $k$ potential sharing groups $C_1, \ldots, C_k$ w.r.t. $\mathcal{M}$ \;
\For{$i \gets \{1, 2, \ldots, k\}$}
{
    $M_{h,j} \gets \mathcal{M}(A_h, A_j)$ \footnotesize{(\emph{Gram matrix over} $Agents(C_i)$)}\;
    $P_i(h,j) \gets \frac{\exp \left(M^i_{h,j} \right)}{  \sum_{a, b \in C_i} \exp \left( M^i_{a, b}\right)}$ \footnotesize{(\emph{sampling distribution})}\;
    \For{each agent $a \in C_i$}
    {
        \For{each agent $b \in Agents(C_i) - \{a\}$}
        {
		    With probability $P_i(a,b)$\;
			{
                $\quad$ Let $\Psi(a) \gets \Psi(a) \cup \{b\}$\;
				$\quad$ Let $Agents(C_i) \gets Agents(C_i) - \{b\}$\;
			}
        }
    }
}
\caption{Selection of Sharing Partners}
\label{alg:sharing-partners}
\end{algorithm}

The process for selecting sharing partners, described in Algorithm \ref{alg:sharing-partners}, is repeated (in parallel) by each supervisor once every $K$ steps. Let $n$ be the number of agents in a supervisory group (which can be defined to include only a small and bounded fraction of the total number of agents in the system) and $d$ be the dimensionality of the context feature vector computed by $f$. Under mild assumptions\footnote{e.g., that the number of iterations executed by the clustering algorithm is  proportional to $n$ and that the complexity of computing $f$ is proportional to the number of observations used to construct such features.} it is possible to show that the complexity of Algorithm \ref{alg:sharing-partners} is $O(dn^3)$. In practical terms, the time-complexity is dominated by the inversion of an $(n \times n)$ matrix, needed in order to compute distances according to the metric proposed in Section \ref{sec:context-based-learning}. If other distance metrics are used (e.g., Euclidean distances) the complexity of the method becomes quadratic in $n$ and linear in $d$. Importantly, notice that because this process is executed separately and independently by each supervisor, the overall complexity of the process is independent of the number of supervisors in the system---it depends crucially only on the number of agents being overseen by each supervisor. The communication complexity of the method is linear in the number $k$ of potential sharing groups, linear in the number of agents in each potential sharing group (i.e., at most $n$) and linear in the reporting interval $K$: $O(kKn)$. Empirically, the communication costs of Algorithm \ref{alg:sharing-partners} seem to scale linearly with the supervisor-to-subordinate ratio (see Section \ref{sec:scalability} for more details).

% PARA CADA SUPERVISOR:
% 	com n subordinates
% 	K reporting time
%
% calcular V:
%    n * K * complexidade_de_f(dadas <=K amostras)
%
% rodar k-means sobre V
%    O(i*k*n*d)
%      i iteracoes
% 	 k clusters
% 	 n pontos/vetors
% 	 d dimensoes
%
% construir gram matrix:
%    inverte matriz de covariancia 1 vez, com O(n^3)
%    e depois calcula n*n distancias
%
% depois o processo eh sao 2 for's encadeados ate n
%
% n*K*complexidade_f + iknd + n^3 + n^2 + n^2
%
% como K<n,
%
% n*n*complexidade_f + innd + n^3 + n^2 + n^2
% =
% n^2 * complexidade_f + n^2 * (i*d) + n^3
%
% n^3 + n^2 (complexidade_f + id)
%
% if we bound i (#iters of k-means by n, the number of agents)
%
% n^3 + n^2 (complexidade_f + nd)
% = n^3 + n^3 d + n^2 complexidade_f
% = n^3 d + n^2 complexidade_f
% se complexidade_f dadas <K amostras (e como K<N) for linear em K (e portanto linear em n),
% = n^3 d + n^2 n
% = n^3 d

\section{Experiments}
\label{sec:results}

% BrunoDelete
% \begin{figure*}
% \centering
% \begin{subfigure}{0.22\linewidth}
% \centering
% \includegraphics[width=\linewidth]{figures/boundary-distribution.pdf}
% \caption{Border Task Concentration}
% \end{subfigure}
% \qquad
% \begin{subfigure}{0.22\linewidth}
% \centering
% \includegraphics[width=\linewidth]{figures/center-distribution.pdf}
% \caption{Center Task Concentration}
% \end{subfigure}
% \caption{Task concentration patterns explored in this work. Agents having a non-zero environmental arrival rate are shaded in red.}
% \label{fig:task-concentration}
% \end{figure*}
% BrunoAdded
% \begin{figure}[!ht]
% \centering
% \includegraphics[width=\linewidth]{figures/patterns_with_subcaptions.pdf}
% %\caption{(a) Border Task Concentration (b) Center Task Concentration.}
% \caption{Task concentration patterns explored in this work. Agents having a non-zero environmental arrival rate are shaded in red.}
% \label{fig:task-concentration}
% \end{figure}

We evaluate our algorithm on large network-distributed task allocation problems (Figure \ref{fig:task-allocation-example}). An agent maintains two queues of tasks: a \emph{processing queue}, with tasks that it has committed to work on; and a \emph{routing queue}, with tasks that are not actively being worked on and that can be forwarded to a neighbor or processed locally. Each task has a duration $s$. %,  indicating how many time units it takes to complete it. %After a task has been at the head of an agent's processing queue for $s$ steps, it is dequeued and marked as completed.  
 The reward function is defined as the reciprocal of the average service time over a time window; service time is the time incurred from task creation to completion. In all experiments, task duration is an exponentially-distributed random variable with mean 10. Tasks are generated by the environment according to patterns that are \emph{unknown} to the agents, which (along with the fact that agents cannot observe their neighbors' states and policies) makes the problem non-stationary. When a task is created, it is associated with some agent $v$ and placed in its routing queue. Upon executing an action (to either process or forward a task) agents receive a reward of $\frac{1}{d}$, where $d$ is the estimated service time of the agent receiving the task. To estimate service time, agents keep track of the time taken to complete past tasks. Agents learn policies using an extension of Q-Learning to the multi-agent case with stochastic policies, which is known to outperform related methods in domains similar to ours \cite{zhang2010multi}. %we investigate here.
%Agents learn policies using an extension of Q-Learning to the multi-agent case with stochastic policies, which is guaranteed to converge to a Nash equilibrium \cite{zhang2010multi}.% to identify policies that minimize average service time. 

% Agent $v$ has several options available for dealing with the tasks in its routing queue; it may either work on the task itself, by adding the task to its processing queue, or it may forward the task to a neighbor directly connected to it. An agent may forward a task to any agent that it is directly connected to; that is, for all $j \in V$, such that $(v, j) \in E$, agent $v$ has action \texttt{forward-j} available to it, which adds a task to the routing queue of agent $j$. 

In our experiments, context features for agent $i$ are composed of three quantities: $i$'s load relative to the mean load of its neighbors, and the rate at which each of its neighbors receives tasks from the environment and from other agents. Since agents with different neighborhood sizes have different actions spaces, their observations have different dimensionality; we therefore restrict context comparisons to agents with the same number of neighbors. Experience sharing between agents with different action spaces is beyond the scope of this paper and would require learning inter-task mappings (e.g. see \cite{taylor2007}). Supervisory groups in these experiments were defined according to the criterion discussed in Section \ref{sec:performance}. %are assigned according to the network quadrant their supervisor belongs to. 
When applying Algorithm \ref{alg:sharing-partners}, we used the $K$-means algorithm with Mahalanobis distance. We automatically set the number $k$ of clusters based on the gap statistic \cite{tibshirani2001estimating}. The task-allocation networks used in our experiments are lattices of up to 729 agents, where each agent directly interacts with 4 neighbors. Different network instances were obtained by varying two parameters regulating the type of task distribution to be tackled by the agents in the system:

\begin{itemize}
    \item \textbf{Task Concentration/Pattern}: %Either \textit{border} or \textit{center}. 
	This parameter regulates whether tasks originate at the \emph{outer} edges of the network or at \emph{central} nodes. Each pattern requires a qualitatively different system behavior. A policy for the border concentration requires boundary agents to forward tasks inward, and central agents to accept tasks; a policy for the center concentration requires the opposite arrangement;
    \item \textbf{Task Frequency}: Tasks are generated with frequency governed by a Poisson distribution. For agents that do not receive tasks from the environment, $\lambda=0$. For all others, a fixed $\lambda > 0$ is used. We consider a set of 11 $\lambda$ values selected uniformly along the range $[0.25, 0.35]$. We do not consider $\lambda<0.25$  since even random policies perform well in this case, nor $\lambda>0.35$, since this leads to queues that grow indefinitely even under optimal policies.
	
\end{itemize}

% (i.e., are neither border not central nodes)
%These patterns are shown in Figure~\ref{fig:task-concentration}. 

% \begin{enumerate}[label=(\alph*),noitemsep,topsep=0pt,parsep=0pt,partopsep=0pt]
%If $i$ has 4 neighboring agents, for instance, its context feature vector has length 9 (four components of type \emph{(a)}, four components of type \emph{(b)}, and one component of type \emph{(c)}). 

\subsection{Performance under Experience Sharing}
\label{sec:performance}

We first examine the impact of the number of supervisors on system performance. On one hand, a single-supervisor configuration results in a nearly centralized system which benefits significantly from sharing opportunities. This corresponds to the case where all agents are placed in a same potential sharing group and \emph{may} undergo experience sharing. Alternatively, no supervisors could exist, in which case the system corresponds to a conventional MAS with no information sharing. Intermediate sharing configurations are possible, with different numbers of supervisors and corresponding subordinate agents. Note that the single-supervisor configuration is often infeasible in real environments, as it is burdened with high communication costs (see Section \ref{sec:scalability}). Four alternative supervisory structures were considered in our experiments. First, we evaluated \emph{two baseline} configurations: one with no supervision, corresponding to a conventional MAS with no information sharing; and one with a single supervisor, corresponding to a system where \emph{all} agents may share information. Intermediate sharing configurations, with 4 and 9 supervisors, were also investigated. The single-supervisor configuration has a supervisor-subordinate ratio of 1:99, the 4-supervisor configuration 1:24, and the 9-supervisor configuration roughly 1:10. Subordinate agents were assigned to supervisors in a way that minimizes the network distance between pairs of subordinates. Results discussed in this section correspond to $440$ runs of our algorithm; in particular, we executed five trials of each combination of task concentration pattern, value of $\lambda$, and supervisory structure were performed, for a total of $2 \times 11 \times 4 \times 5 = 440$ runs.

\begin{figure}[!ht]
    \centering
    \includegraphics[width=0.95\linewidth]{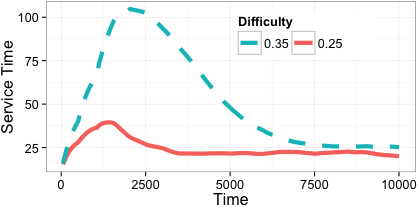}
    \caption{Performance under different difficulty settings $\lambda$.}
    \label{fig:difficulty-comparison}
\end{figure}

To appreciate the difference between the least challenging task allocation setting ($\lambda=0.25$) and the most challenging one ($\lambda=0.35$), we analyze the average system-wide service time obtained by the \emph{single-supervisor} configuration throughout 10,000 steps, with tasks concentrated on the border of a 100-agent lattice. Figure~\ref{fig:difficulty-comparison} shows the evolution of service time as time progresses. At first, poor policies lead to a heavily saturated system, which degrades service times, with a peak of approximately 100 steps per task occurring about 25\% of the way into the simulation. As agents learn appropriate policies, they more rapidly complete tasks, ultimately converging to a service time of about 25 steps per task. This level of performance is reached regardless of $\lambda$, though the amount of time taken to reach it, and the performance of the system during learning, are both of central importance.

\begin{figure}[!ht]
\centering
\includegraphics[width=0.95\linewidth]{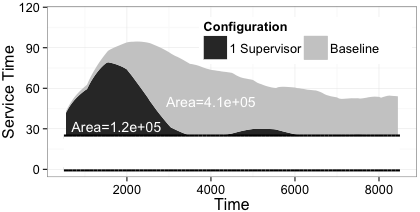}
\caption{Performance of the single-supervisor configuration vs. no-sharing. Smaller areas under the curve indicate faster convergence.}
\label{fig:performance-example}
\end{figure}

\begin{figure}[!ht]
\centering
\includegraphics[width=0.95\linewidth]{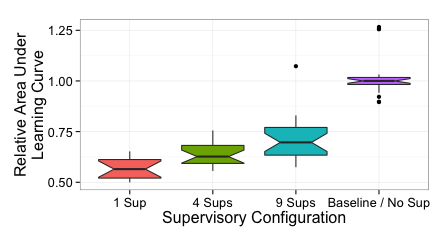}
\caption{Performance of different supervisory configurations in a 100-agent network; smaller values correspond to faster learning.}
\label{fig:relative-performance}
\end{figure}

In all experiments that follow we define performance as the area under the curve of service time as a function of time. When the system converges quickly, this area is small. We treat the minimum service time ever attained by any configuration as zero, so that running the system at the optimal performance does not accumulate area; i.e., performance of an optimally-performing system is \emph{invariant} with respect to simulation duration (see Figure~\ref{fig:performance-example} for an example). 
Figure %~\ref{fig:combined}a 
\ref{fig:relative-performance} shows that the single-supervisor configuration far outperforms the baseline approach with no transfer, with information-sharing agents accumulating nearly \emph{half} the area under the curve compared to agents that do not share experiences. As additional supervisors are introduced, this benefit diminishes, since there are fewer experience sharing opportunities within each supervisory group. Note, however, that even with a high supervisor-subordinate ratio of 1:10 (which corresponds, in this experiment, to having approximately as many supervisors as agents in each supervisory group), experience sharing still allows us to reduce the learning curve area by \emph{more than 25\%}.

% We then normalized this performance measure relative to the median area of all trials of the corresponding baseline configuration, allowing summarization of performance results across settings of task concentration and frequency.

\subsection{Scalability and Communication Overhead}
\label{sec:scalability}

% \begin{figure*}[!ht]
% \centering
% \includegraphics[width=0.9\textwidth]{figures/combined_figs.png}
% \caption{\textbf{(a)} Performance of different supervisory configurations in a 100-agent network; smaller values correspond to faster learning. \textbf{(b)} Performance of the 9-supervisor configuration as a function of network size.}
% \label{fig:combined}
% \end{figure*}

\begin{figure}[!ht]
\centering
\includegraphics[width=0.95\linewidth]{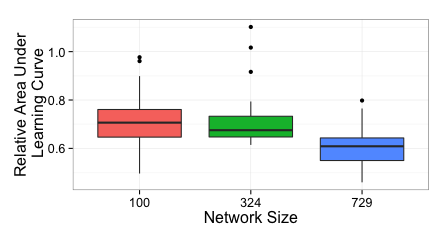}
\caption{Performance of the 9-supervisor configuration as a function of network size.}
\label{fig:vary-network-size}
\end{figure}

We intuitively expect that information sharing becomes more beneficial as the size of the system grows: larger systems typically have a more diverse pools of agents which may benefit from sharing. To test this hypothesis, we constructed simulations sweeping across a large number of settings for task concentration and frequency, and varied the number of agents through $\{100, 324, 729\}$ (i.e., lattices of dimension 10, 18, and 27). Two supervisory configurations were considered: a 9-supervisor configuration and a baseline (or no-sharing) arrangement. Our goal is to characterize how a 9-supervisor setting fares compared to the baseline as the number of subordinates per supervisor increases. This was achieved by varying the network size (see Figure~%\ref{fig:combined}b)
\ref{fig:vary-network-size}). Performance in the 100-agent network was roughly \emph{30\% higher} than the baseline. As network size increased to 729 agents, performance median \emph{improved by 40\%} compare to the baseline.

These gains come at the cost of increased communication. Note, however, that the communication overhead of Algorithm \ref{alg:sharing-partners} scales with the supervisor-to-subordinate ratio, not with the total number of agents (see Section \ref{subsection:context-similarity} for a formal complexity analysis); e.g., the 9-supervisor configuration undergoes 9 times less communication than the single-supervisor configuration. In our experiments we further observed that communication volume was \emph{invariant} with respect to $K$: on average 43 bytes per step per subordinate using a loss-less compression scheme. This suggests \emph{1)} that communication costs (i.e.,  the total amount of bytes exchanged between agents sharing experiences) scales \emph{linearly} with the supervisor-to-subordinate ratio; and \emph{2)} that even when accounting for communication costs, more distributed configurations tend to perform better. In fact, all evaluated information-sharing configurations surpassed the baselines while incurring very low communication overhead---as previously mentioned, on average 43 bytes per step per subordinate using a loss-less compression scheme in a 100-agent network. The fact that communication volume was empirically observed to be invariant with respect to $K$ is not a trivial statement---in the worst case, communication volume could still increase linearly with the number of potential sharing groups within a supervisory group, $k$, and linearly with the reporting window, $K$. The fact that it does not suggests that Algorithm \ref{alg:sharing-partners} is capable of effectively identifying and exploiting useful sharing opportunities, instead of always relaying all $K$ observations to all $n$ agents within a supervisory group.
		
\begin{figure}[!ht]
\centering
\includegraphics[width=\linewidth]{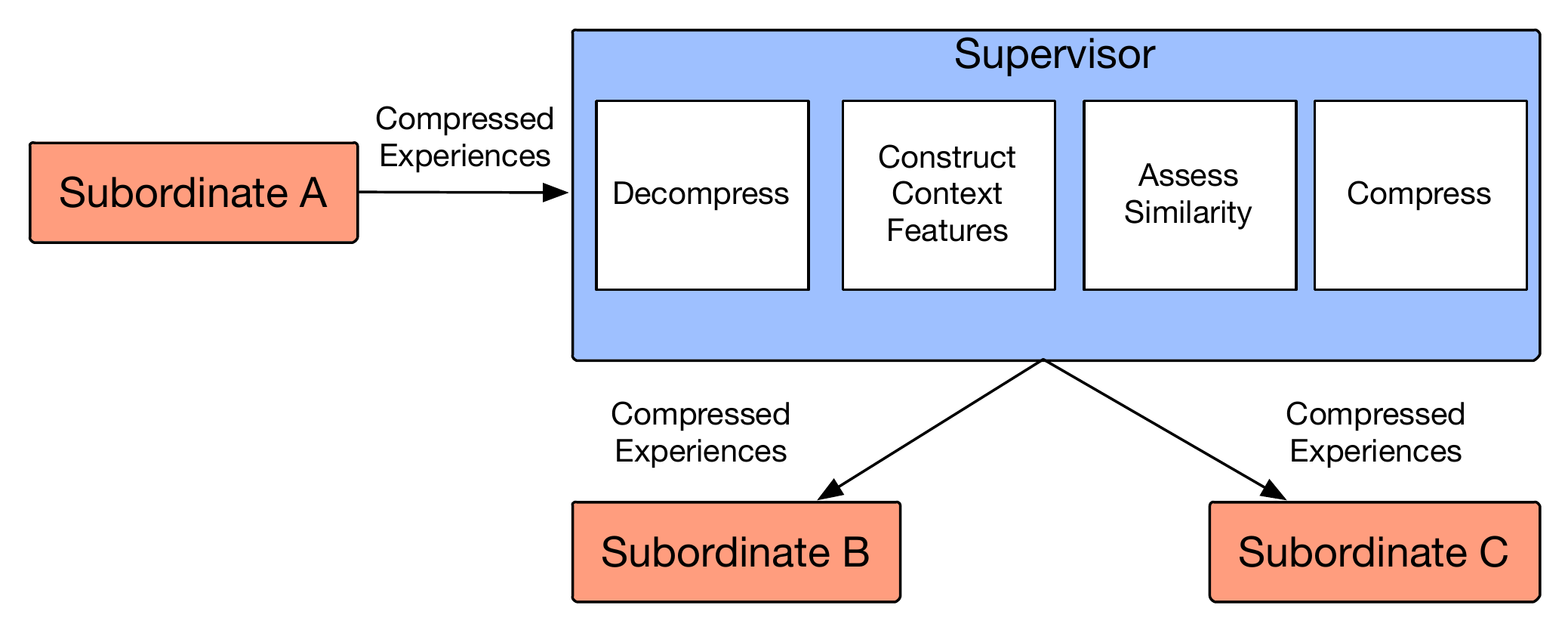}
\caption{Relaying compressed experiences through a supervisor.}
\label{fig:supervisor-relay}
\end{figure}

We also explored the use of lossy experience compression schemes, which significantly reduced communication costs and incurred negligible performance penalties. One lossy compression technique that we evaluated is a sparse polynomial spline interpolator---a method that approximately represents a set of experiences with as few coefficients as possible. Supervisors may use such a sparse interpolator in order to model how the observed data (i.e., sequences of states, actions, and rewards within a reporting window) vary with time. Because states and rewards usually change smoothly, the number of coefficients needed to represent the corresponding set of observations $O_i$ is typically much smaller than the number of observations ($K$). Note that actions within the set of observations $O_i$ of an agent $i$ are categorical features, and therefore are not compressed. We constructed each compressed model of $O_i$ according to different \emph{compression degrees}. Compression degree refers to the frequency with which we subsample elements of $O_i$ in order to construct the training set for the interpolator. A compression degree of $R$ typically results in models requiring $O(\frac{K}{R})$ coefficients in order to approximate a set of $K$ observations. When employing a lossy experience compression scheme such as this, supervisors relay not complete sets of experiences to all agents within a sharing group, but only the coefficients of the corresponding lossy model. Figure \ref{fig:supervisor-relay} depicts how the use of a compression scheme impacts the communication framework by which subordinate and supervisor agents in a network share experiences. 

\begin{figure}[!ht]
\centering
\includegraphics[width=0.9\linewidth]{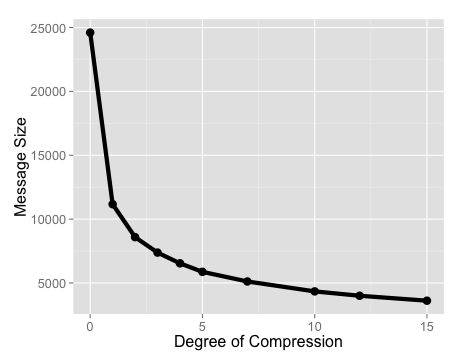}
\caption{System-wide communication volume resulting from the use of different lossy compression degrees.}
\label{fig:compression}
\end{figure}

Figure \ref{fig:compression} shows the system-wide communication volume (in bytes) resulting from the use of different compression degrees. In particular, this graph presents the average system-wide communication volume when evaluated over all supervisory structures discussed in Section \ref{sec:performance} and tested in a network with 100 agents. The reporting interval in this experiment was $K=100$. When performing these experiments observed an interesting trend: the use of lossy models with compression degree up to $15$ had negligible effect on the performance of the method (smaller than error bars in Figure \ref{fig:relative-performance}). This occurs because the set of observations of an agent (states and rewards) is highly temporally correlated, and can, therefore, be efficiently compressed via a sparse model and reconstructed with very little information loss. Compression degrees higher than $15$, on the other hand, resulted in negligible positive impact on the overall system-wide communication volume, since the size of the (uncompressed) action time series begins to dominate. These observations suggest that in systems where states and rewards vary smoothly over time, it is possible to deploy effective compression schemes for lowering the overall communication costs of the method---in this application, resulting in a 5-fold decrease when compared to an architecture that uses loss-less compression.

\subsection{Robustness}

\begin{figure}[!ht]
\centering
\includegraphics[width=1.0\linewidth]{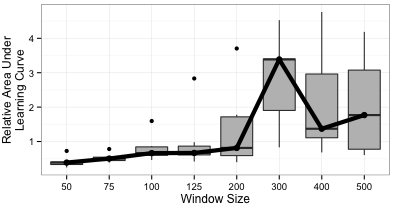}
\caption{Effect of varying reporting intervals $K$ on system performance.}
\label{fig:vary-window-size}
\end{figure}

In the previous experiments, unless noted otherwise, we used a reporting interval of $K=115$ steps, selected by cross-validation to minimize a balance between performance and communication overhead. Smaller values of $K$ lead to more frequent communication, whereas larger values of $K$ decrease the likelihood that an agent's transition and reward models will remain static across the $K$-timestep interval. The latter case results in reports containing mixed observations arising from multiple underlying local learning contexts, which makes sharing less effective. We evaluate the robustness of our algorithm by studying the effect of using suboptimal reporting intervals $K$. We ran 10 trials of the single-supervisor and baseline configurations for each of eight reporting intervals, using $\lambda=0.3$ in a 100-agent network with boundary-based task distribution (Figure~\ref{fig:vary-window-size}). As larger reporting intervals are used, performance degrades, as heterogeneity is induced in transition and reward samples and the agent learns a policy that averages observations from different local learning contexts.
% TODO: 5) Re:peak in Fig7. Performance drops at K=300 because agents transition to a new context for which the observation history is still undersampled. We will add a discussion.

\begin{figure}[!ht]
\centering
    \includegraphics[width=1.0\linewidth]{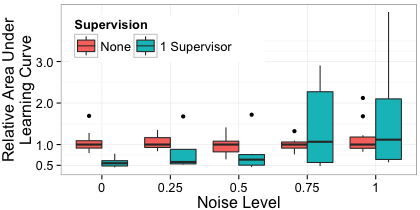}
\caption{Robustness of experience sharing to the use of sub-optimal/corrupted context features.}
\label{fig:context-feature-noise}
\end{figure}

We also analyzed our method's robustness by studying the impact of using corrupted or suboptimal context features. Context features that do not properly abstract the underlying local learning environment make it difficult to identify appropriate sharing opportunities. To evaluate the sensitivity of our algorithm to this issue we added different levels of normally-distributed noise to context features. Noise degrades the quality of the signal encoded in the features, up to a point where they are entirely uncorrelated with the underlying system dynamics. The magnitude of the noise was varied relative to the standard deviation of context features; when noise level is 1, the standard deviation of the normally-distributed noise term is greater than (or equal) to the standard deviation of any context feature, effectively eliminating any signal that they encoded. Figure~\ref{fig:context-feature-noise} shows that when noise dominates (approaches 1), performance becomes increasingly volatile. The performance distribution, with mean approximately 1, suggests that as context features become less meaningful, the sharing mechanism is equally likely to achieve a 50\% reduction in the area under the learning curve as it is to increase this area by 100\%. In other words, as the information-sharing process tends to be guided by biased or incorrect features, there is no consistent positive or negative impact on performance; the most prominent impact is on performance \emph{variability}.
% "arising from multiple underlying... , i.e., in situations where the agent perceives clearly different observations (of state transition and rewards) due to changing policies of its neighbors"
% standard deviation of the normally-distributed noise term is greater than or equal to the standard deviation of any context feature, 

\section{Discussion}

We have presented a solution for experience transfer among RL agents in large multi-agent systems. Our method adaptively identifies opportunities to transfer experiences between context-compatible agents, where contexts provide abstract characterizations of local learning environments. By explicitly identifying context-compatible groups, we avoid issues arising from the use of policy, Q-function or model similarity as proxies of environment similarity. Although other transfer mechanisms are possible, to our knowledge no other methods exist that address our particular setting and scale. We believe this is the first algorithm that allows experience sharing in a concurrent and interacting MAS with \mytilde1000 agents while undergoing low communication and computational overhead. Importantly, the time complexity of our method scales with the number of agents within each local supervisory group, not with the total number of agents in the network. Experiments further suggest that the method provides significant improvements over baseline settings with no experience sharing, and quantitative analyses demonstrate that sharing becomes increasingly advantageous as the system size grows. Finally, we have shown that our method is robust to noise-corrupted or suboptimal context features, that communication costs scale linearly with the supervisor-to-subordinate ratio, and that sparse lossy compression schemes may be deployed and provide a 5-fold improvement in communication costs while inducing negligible negative impact on system-wide performance.

% original:
% Although other transfer mechanisms may be possible, to our knowledge no other methods have been proposed that address this setting and scale. We believe this is the first algorithm that allows experience transfer in a concurrent and interacting multi-agent environment with low communication and computational overhead. Our experiments suggest that the method provides significant improvements over a baseline learning setting wit no experience sharing. Quantitative analyses demonstrate that information sharing becomes increasingly advantageous as the system size grows. Finally, we have shown that our method is robust to noise-corrupted or suboptimal context features, and that communication costs scale linearly with the supervisor-to-subordinate ratio.

%As future work, we are mainly interested in mechanisms for context feature selection. For model-free learning algorithms, this could require a way of comparing context feature distances with a measure of divergence between true state transition and reward distributions.

 % Finally, we note that in model-base settings it may be possible to bypass the explicit need for context features entirely, by designing an online measure of estimated transition and reward divergence. This could be advantageous whenever \emph{a priori} and explicit knowledge about the underlying system dynamics is available

\bibliographystyle{abbrv}
\bibliography{ms}  % sigproc.bib is the name of the Bibliography in this case

\begin{thebibliography}{10}

\bibitem{boutsioukis2012transfer}
G.~Boutsioukis, I.~Partalas, and I.~Vlahavas.
\newblock Transfer learning in multi-agent reinforcement learning domains.
\newblock In {\em Recent Advances in Reinforcement Learning}, pages 249--260.
  Springer, 2012.

\bibitem{carroll2005task}
J.~L. Carroll and K.~Seppi.
\newblock Task similarity measures for transfer in reinforcement learning task
  libraries.
\newblock In {\em Proceedings of the International Joint Conference on Neural
  Networks}, pages 803--808. IEEE, 2005.

\bibitem{de2000functional}
R.~M. De~Jong and J.~Davidson.
\newblock The functional central limit theorem and weak convergence to
  stochastic integrals i: weakly dependent processes.
\newblock {\em Econometric Theory}, pages 621--642, 2000.

\bibitem{dietterich2000}
T.~G. Dietterich.
\newblock State abstraction in maxq hierarchical reinforcement learning.
\newblock In {\em Advances in Neural Information Processing Systems 12}, pages
  994--1000. MIT Press, 2000.

\bibitem{garant2015accelerating}
D.~Garant, B.~C. da~Silva, V.~Lesser, and C.~Zhang.
\newblock Accelerating multi-agent reinforcement learning with dynamic
  co-learning.
\newblock Technical Report UM-CS-2015-004, School of Computer Science,
  University of Massachusetts Amherst, 2015.

\bibitem{guestrin2001multiagent}
C.~Guestrin, D.~Koller, and R.~Parr.
\newblock Multiagent planning with factored mdps.
\newblock In {\em Proceedings of Advances in Neural Information Processing
  Systems}, pages 1523--1530, 2001.

\bibitem{hu2015learning}
Y.~Hu, Y.~Gao, and B.~An.
\newblock Learning in multi-agent systems with sparse interactions by knowledge
  transfer and game abstraction.
\newblock In {\em Proceedings of the International Conference on Autonomous
  Agents and Multiagent Systems}, pages 753--761. IFAAMAS, 2015.

\bibitem{keller2006}
P.~W. Keller, S.~Mannor, and D.~Precup.
\newblock Automatic basis function construction for approximate dynamic
  programming and reinforcement learning.
\newblock In {\em Proceedings of the 23rd International Conference on Machine
  Learning}, pages 449--456, New York, NY, USA, 2006. ACM.

\bibitem{kitano1999robocup}
H.~Kitano, S.~Tadokoro, I.~Noda, H.~Matsubara, T.~Takahashi, A.~Shinjou, and
  S.~Shimada.
\newblock Robocup rescue: Search and rescue in large-scale disasters as a
  domain for autonomous agents research.
\newblock In {\em Systems, Man, and Cybernetics, 1999. IEEE SMC'99 Conference
  Proceedings. 1999 IEEE International Conference on}, volume~6, pages
  739--743. IEEE, 1999.

\bibitem{konidaris2009}
G.~Konidaris and A.~G. Barto.
\newblock Efficient skill learning using abstraction selection.
\newblock In {\em Proceedings of the 21st International Joint Conference on
  Artificial Intelligence}, pages 1107--1112, 2009.

\bibitem{kretchmar2002}
R.~M. Kretchmar.
\newblock Parallel reinforcement learning.
\newblock In {\em Proceedings of the 6th World Conference on Systemics,
  Cybernetics, and Informatics}, 2002.

\bibitem{lazaric2008transfer}
A.~Lazaric, M.~Restelli, and A.~Bonarini.
\newblock Transfer of samples in batch reinforcement learning.
\newblock In {\em Proceedings of the International Conference on Machine
  Learning}, pages 544--551. ACM, 2008.

\bibitem{li2006}
L.~Li, T.~J. Walsh, and M.~L. Littman.
\newblock Towards a unified theory of state abstraction for mdps.
\newblock In {\em Proceedings of the Ninth International Symposium on
  Artificial Intelligence and Mathematics}, pages 531--539, 2006.

\bibitem{littman2001value}
M.~L. Littman.
\newblock Value-function reinforcement learning in markov games.
\newblock {\em Cognitive Systems Research}, 2(1):55--66, 2001.

\bibitem{mahadevan2005}
S.~Mahadevan.
\newblock Proto-value functions: Developmental reinforcement learning.
\newblock In {\em Proceedings of the 22nd International Conference on Machine
  Learning}, pages 553--560, New York, NY, USA, 2005. ACM.

\bibitem{mnih2016}
V.~Mnih, A.~P. Badia, M.~Mirza, A.~Graves, T.~P. Lillicrap, T.~Harley,
  D.~Silver, and K.~Kavukcuoglu.
\newblock Asynchronous methods for deep reinforcement learning.
\newblock {\em CoRR}, abs/1602.01783, 2016.

\bibitem{murphy2012machine}
K.~P. Murphy.
\newblock {\em Machine learning: a probabilistic perspective}.
\newblock MIT press, 2012.

\bibitem{nair2005networked}
R.~Nair, P.~Varakantham, M.~Tambe, and M.~Yokoo.
\newblock Networked distributed pomdps: A synthesis of distributed constraint
  optimization and pomdps.
\newblock In {\em Proceedings of the National Conference on Artificial
  Intelligence}, volume~5, pages 133--139, 2005.

\bibitem{oliehoek2008exploiting}
F.~A. Oliehoek, M.~T. Spaan, S.~Whiteson, and N.~Vlassis.
\newblock Exploiting locality of interaction in factored dec-pomdps.
\newblock In {\em Proceedings of the 7th international joint conference on
  Autonomous agents and multiagent systems-Volume 1}, pages 517--524.
  International Foundation for Autonomous Agents and Multiagent Systems, 2008.

\bibitem{owen1982}
G.~Owen.
\newblock {\em Game Theory}.
\newblock Academic Press, 2nd edition, 1982.

\bibitem{paar2007}
R.~Parr, C.~Painter-Wakefield, L.~Li, and M.~Littman.
\newblock Analyzing feature generation for value-function approximation.
\newblock In {\em Proceedings of the Twenty-Fourth International Conference on
  Machine Learning}, page 737–744, 2007.

\bibitem{price2003accelerating}
B.~Price and C.~Boutilier.
\newblock Accelerating reinforcement learning through implicit imitation.
\newblock {\em Journal of Artificial Intelligence Research}, 19:569--629, 2003.

\bibitem{balaraman2003}
B.~Ravindran and A.~G. Barto.
\newblock {SMDP} homomorphisms: An algebraic approach to abstraction in
  semi-markov decision processes.
\newblock In {\em Proceedings of the Eighteenth International Joint Conference
  on Artificial Intelligence}, pages 1011--1018, 2003.

\bibitem{simon1996sciences}
H.~A. Simon.
\newblock {\em The sciences of the artificial}, volume 136.
\newblock 1996.

\bibitem{taylor2013transfer}
A.~Taylor, I.~Duparic, E.~Galv{\'a}n-L{\'o}pez, S.~Clarke, and V.~Cahill.
\newblock Transfer learning in multi-agent systems through parallel transfer.
\newblock In {\em Theoretically Grounded Transfer Learning at the International
  Conference on Machine Learning}, 2013.

\bibitem{taylor2009transfer}
M.~E. Taylor and P.~Stone.
\newblock Transfer learning for reinforcement learning domains: A survey.
\newblock {\em The Journal of Machine Learning Research}, 10:1633--1685, 2009.

\bibitem{taylor2007}
M.~E. Taylor, S.~Whiteson, and P.~Stone.
\newblock Transfer via inter-task mappings in policy search reinforcement
  learning.
\newblock In {\em Proceedings of the 6th International Joint Conference on
  Autonomous Agents and Multiagent Systems}, May 2007.

\bibitem{tibshirani2001estimating}
R.~Tibshirani, G.~Walther, and T.~Hastie.
\newblock Estimating the number of clusters in a data set via the gap
  statistic.
\newblock {\em Journal of the Royal Statistical Society: Series B (Statistical
  Methodology)}, 63(2):411--423, 2001.

\bibitem{witwicki2010influence}
S.~J. Witwicki and E.~H. Durfee.
\newblock Influence-based policy abstraction for weakly-coupled dec-pomdps.
\newblock In {\em Proceedings of the International Conference on Automated
  Planning and Scheduling}, pages 185--192, 2010.

\bibitem{zhang2010multi}
C.~Zhang and V.~Lesser.
\newblock {Multi-Agent Learning with Policy Prediction}.
\newblock In {\em Proceedings of the 24th AAAI Conference on Artificial
  Intelligence}, pages 927--934, Atlanta, 2010.

\bibitem{zhang2013coordinating}
C.~Zhang and V.~Lesser.
\newblock {Coordinating Multi-Agent Reinforcement Learning with Limited
  Communication}.
\newblock In {\em Proceedings of the 12th International Conference on
  Autonomous Agents and Multiagent Systems}, pages 1101--1108. IFAAMAS, 2013.

\bibitem{zhang2010self}
C.~Zhang, V.~Lesser, and S.~Abdallah.
\newblock {Self-Organization for Coordinating Decentralized Reinforcement
  Learning}.
\newblock In {\em Proceedings of the 9th International Conference on Autonomous
  Agents and Multiagent Systems}, pages 739--746, 2010.

\end{thebibliography}

% TODO:
% create the ~.bbl file.  Insert that ~.bbl file into the .tex source file and comment out the command \texttt{{\char'134}thebibliography}.
% TODO: 
\balancecolumns % GM June 2007

\end{document}